\documentclass[preprint]{aastex}

\begin{document}

\title{Multipole Analysis of Kicks in Collision of Spinning Binary Black Holes}

\author{Sarah H. Miller}
\affil{Center for Relativity and Department of Physics\\
The University of Texas at Austin, Austin, TX 78712}
\author{Richard A. Matzner}
\affil{Center for Relativity and Department of Physics\\
The University of Texas at Austin, Austin, TX 78712}

\begin{abstract}

Thorne and Kidder give expressions which allow for analytical estimates of the 
``kick", {\it i.e.} the recoil, produced from asymmetrical gravitational 
radiation during the interaction of black holes, or in fact any gravitating
compact bodies. (The Thorne-Kidder formula uses momentum flux calculations based on the
linearized General Relativity of gravitational radiation.)
We specifically treat kicks arising in the binary interaction 
of equal mass black holes, when at least one of the black holes has significant 
spin, \emph{$a$}. Such configurations can produce very large kicks in computational 
simulations. We consider both {\it fly-by} and {\it quasicircular} orbits.
For fly-by orbits we find substantial kicks from those Thorne-Kidder terms which are 
linear in \emph{$a$}. For the quasi-circular case, we consider in addition
the nonlinear contribution ($ O(a^2)$) to the kicks, and provide a dynamical 
explanation for such terms discovered and displayed by 
\citet{2007arXiv0712.2819B}. However, in the cases of maximal kick velocities, 
the dependence on spin is largely linear (reproduced in numerical results 
\citet{2007ApJ...661..430H}). 

\end{abstract}


\section{Introduction}

Gravity waves (gravitational radiation) are a product of an extreme
gravitational environment; the waves propagate through spacetime 
itself as fluctuations in the gravitational field (the curvature of 
spacetime). Experiments, supported by large collaborations such 
as LIGO \citep{1996qpct.conf..101T}, now seek to detect these waves of 
gravity by measuring these
subtle fluctuations, further motivating theorists to 
understand gravity wave emission. Not only is it important to 
characterize these waves in the context of gravitational wave 
observations, but also for cosmological models in which black 
hole mergers play an increasingly important role 
\citep{diMa..2008,Whit..2008}. 

We consider two interacting black holes in a binary system that 
radiate away their energy and angular momentum with gravity waves. 
In addition to 
the orbital angular momentum of the binary, the individual black 
holes can also have their own spin angular momenta.
If the black holes are gravitationally bound their radiation 
will eventually lead to inspiral and collapse. The final 
coalescence of two black holes releases a huge amount of 
gravitational radiation (up to 10 \% of the total rest 
mass (\citet{matznerEtAl2008})). If any sort of asymmetry is present in 
the binary, {\it e.g.} if the black holes have unequal masses, 
or if the black holes have unequal spin angular momenta or
spin angular momenta unaligned with 
their orbital angular momentum, then the asymmetry will be reflected 
in the gravitational radiation emission of the coalescence, 
resulting in a ``kick'' of the final black hole. In this paper we consider 
the equal mass case of black hole merger where at least one black hole has substantial spin.

Not only can we detect
the gravitation radiation of the merger, the kick itself may propel the
resultant black hole completely out of the center of a galaxy. Recent 
numerical simulations of mergers of spinning black holes 
result in kicks up to 4,000 km/s in quasicircular inspiral
\citep{2007PhRvD..75f4030C} and 10,000km/s in hyperbolic 
encounters \citep{healyEtAl2008}, which could 
easily exceed the escape velocity of a galactic nucleus. \citet{2008arXiv0802.3873S} have 
presented limits on observing this phenomenon since kicked supermassive
black holes may retain a portion of their in-falling matter, which
may produce large flares of energy in a characteristic spectrum. 
Recently, \citet{2008ApJ...678L..81K} have discovered strong 
observational evidence of a recoiling supermassive black hole with
optical emission lines. The black hole appears to have a kick of 
2650 km/s!

\section{Multipole Formula}

Before the modern methods of numerical relativity were 
developed, \citet{RevModPhys.52.299} and \citet{PhysRevD.52.821} developed
a multipole formula which describes the gravitational
radiation kick from dynamical gravitating systems. The formula
is based on derivatives of low-order multipoles of the masses 
and spins of the binary:
\begin{equation}
\frac{dP_i}{dt}=\frac{16}{45}\epsilon_{ijk} I^{(3)}_{jl} H^{(3)}_{kl} + \frac{4}{63} H^{(4)}_{ijk}H^{(3)}_{jk} + \frac{1}{126} \epsilon_{ijk}I^{(4)}_{jlm}H^{(4)}_{klm}.
\label{kickEQ}
\end{equation}

We have included only the terms that depend 
on the spin. Eq (\ref{kickEQ}) is made up of nth time-derivatives ${}^{(n)}$ of 
mass quadrupoles and octupoles, $I_{ij}$ and $I_{ijk}$ respectively, and of
spin quadrupoles and octupoles, $H_{ij}$ and $H_{ijk}$ respectively. While 
the mass quadrupole and octupole are fairly familiar, the spin quadrupoles 
and octupole are less so. We present formulae below for these quantities, 
and a scheme to compute the spin multipoles. We use the physical Kerr parameter
$a=j/m$, where $j$ is the angular momentum and $m$ is the mass of the black 
hole, and we point out that $H_{ij}$ and $H_{ijk}$ are linear in the spin $a$.
Thus in Eq (\ref{kickEQ}) the first and third terms are linear in the 
spin, and the second term is quadratic in the spin.

Because Eq (\ref{kickEQ}) involves time derivatives, it requires knowledge 
of the motion of the black holes. But there is no analytical form for 
two-body motion in General Relativity, so we will use Newtonian physics to 
describe the motion of the interacting black holes. (One effect of
this choice is to confine the motion to the initial orbital plane, $z=0$.) Thus our 
results will be accurate but uninteresting for Newtonian motions (slow 
velocities, large impact parameters or orbital separations). For the 
interesting case of relativistic interactions (velocities near $c$, 
small impact parameters or orbital separations) we will obtain qualitative 
results, which nonetheless lead to estimates of kick ratios for different 
configurations and allow an analytical understanding of the kick process.
For the quasicircular case our work is complementary to that of \cite{2008PhRvD..77d4031}, 
which is an extensive study of non-equal mass mergers with either zero or equal but opposite spins; we 
study {\it equal mass} mergers with arbitrary ratios and directions of spins on the two black holes.

\subsection{Mass Quadrupole and Octupole}

The standard expressions for the mass quadrupoles and octupoles are:
\begin{eqnarray}
I_{ij}&=&\sum~ m[x_ix_j-\frac{1}{3}r^2\delta_{ij}], \nonumber \\
I_{ijk} &=& \sum~ m[x_ix_jx_k - \frac{1}{5}r^2(\delta_{ij}x_{k} 
+ \delta_{jk}x_{i} + \delta_{ki}x_{j})],
\label{massEQ}
\end{eqnarray}
where the sum is over both black holes. We use Latin letters for spatial 
indices, $x,y,z$, and do not distinguish between covariant and contravariant indices. We
treat one black hole at a time. The two-hole result is obtained by adding the 
contribution from each hole. Because in our examples the black holes have 
equal mass, the mass octupole vanishes on this addition, so that the third term in 
the formula Eq (\ref{kickEQ}) can be ignored regardless of spin 
configuration. Since we need the third time derivative 
of the mass quadrupole, we evaluate $x_k(t)$ and its 
first three time derivatives. 

\subsection{Spin Quadrupole and Octupole}

The spin multipoles require the baffling concept of spin density, but since 
we are dealing with Kerr black holes, which have an intrinsic spin dipole 
moment, we use a trick to evaluate the spin quadrupole. We replace the spin 
dipole by a fictitious pair of spin charges, of value $q=\pm a$ separated by a 
distance m, centered at the actual location of the black hole \citep{2007ApJ...661..430H}. 
This reproduces the dipole angular momentum ($am$), and 
allows us to compute the quadrupole directly: 
\begin{eqnarray}
\label{spinEQ}
H_{ij} &=& \sum q~ [x_{i} x_{j} - \frac{1}{3} r^{2}
 \delta_{ij}]\nonumber\\
H_{ijk} &=& \sum q~ [x_{i} x_{j} x_{k} - \frac{1}{5} r^{2}
 (\delta_{ij} x_{k} + \delta_{jk} x_{i} + \delta_{ki} x_{j})]\,.
\end{eqnarray}

Now the sum is over the two black holes and over the two spin charges for each hole,
and the $x_i $ appearing in these formula are offset in the direction of the spin.
Strictly one should take a limit to small separations with the product $am$ held constant, 
but this is automatic here because the result is proportional to $am$.
We can separately consider the cases for spin components, $a_x$, $a_y$, $a_z$, 
and add the spin quadrupole or octupole components after individual calculation. 
(To the lowest order, the spin is parallel-transported along the orbit.)

\section{Hyperbolic Fly-by}

Consider equal mass black holes approaching each other with impact parameter 
$2b$ in the $x$-$y$ plane,  and equal and opposite velocities (in the center 
of mass frame) of $v_0$. We assume the motion starts along the $x-$axis, and 
the impact parameter is so large that the angle of deflection is small. 
Then, rather than explicitly computing the hyperbolic orbit, we can simplify 
the analysis by assuming that the motion is uniform in the $x-$direction. 
We will however compute the acceleration (and higher derivatives) using this 
uniform time dependence. Also, although the radiation reaction can change 
the orbital plane, we will not consider this feedback in the computation 
of the radiation. These points should become clear as we work through the 
analysis. In this section we consider only the lowest, linear-in-$a$ term in 
Eq (\ref{kickEQ}).

Let us consider only the black hole initially moving along the line $y=+b 
=const$ with velocity $v=v_0$. It is at $x=0$ at time $t=0$. The acceleration 
is given by the Newtonian result:
\begin{equation}
\frac{d^2x}{dt^2}=-\frac{mx}{4r^3} = -\frac{mv_0t}{4(b^2+(v_0t)^2)^\frac{3}{2}},
\label{accelx}
\end{equation}
where $2r=2(b^2+(v_0t)^2)^\frac{1}{2}$ is the separation between the black holes.
Similarly,
\begin{equation}
\frac{d^2y}{dt^2}=-\frac{mb}{4r^3} = -\frac{mb}{4(b^2+(v_0t)^2)^\frac{3}{2}}.
\label{accely}
\end{equation}
(Our units have the Newtonian constant $G=1$, and we henceforth take $c=1$ also.)
Integrating Eqs (\ref{accelx}) \& (\ref{accely}) once in time yields:
\begin{equation}
v_x = v_0 +\frac{m}{4v_0(b^2+(v_0t)^2)^\frac{1}{2}},
\label{vx}
\end{equation}
and 
\begin{equation}
v_y = -\frac{mt}{4b(b^2+(v_0t)^2)^{\frac{1}{2}}}.
\label{vy}
\end{equation}
Similarly we can differentiate Eqs (\ref{vx}) \& (\ref{vy}) to obtain higher time 
derivatives of the position. 

In the Newtonian approximation the motion remains in the $x-y$ plane 
($z=0$), and the motion of the equal mass black holes is symmetrical through 
the origin. Thus the contribution to the mass quadrupole is equal for the 
two masses. And, because of the symmetry, only $I_{xx}$, $I_{xy}$, $I_{yy}$, 
and $I_{zz}$ are in principle nonzero. In particular, $I_{xy}=-\sum 
m xy$ and $I_{zz}=-\sum\frac{1}{3}m (x^2+y^2)$. 

Note that the deflection angle is of $O(m/bv_0)$. Consistent 
with our approximation, we keep only the lowest powers of $(m/b)$ in 
computing multipoles. 
Our approach is encapsulated in the following rules: 
\begin{enumerate}\setlength{\itemsep}{-3pt}
\item write the desired derivative of the multipole in terms of derivatives of $x$ and $y$;
\item replace undifferentiated $x$ factors by $v_0t$;
\item replace undifferentiated $y$ factors by $b$;
\item replace $\dot x$ (first time-derivative of $x$) factors by $v_0$;
\item in any term with a product of derivatives, first apply rules (3) and (4) above, 
then drop any term with more than one remaining differentiated factor.
\end{enumerate}
This approach is similar to those of \cite{1989PThPh..82..535O} and \cite{1990MNRAS.242..289B}.

We demonstrate the approach by evaluating $I^{(3)}_{xy}$ using our prescription.
We treat only one hole (the one with $v_x \approx +v_0$):
\begin{eqnarray}
I^{(3)}_{xy}&=&m(y\frac{d^3x}{dt^3} +3\frac{dy}{dt}\frac{d^2x}{dt^2}+3\frac{d^2y}{dt^2}\frac{dx}{dt}+\frac{d^3y}{dt^3}x) \nonumber \\
  &\rightarrow& m(b\frac{d^3x}{dt^3} +3\frac{d^2y}{dt^2}v_0+\frac{d^3y}{dt^3}v_0t) \nonumber \\
 &=& \frac {m^2bv_0((v_0t)^2 - 2b^2)}{2((v_0t)^2 + b^2)^{\frac{5}{2}}}.
\end{eqnarray}
As another example, we explicitly evaluate $I^{(3)}_{zz}$:
\begin{eqnarray}
I^{(3)}_{zz}&=&-\frac{2}{3}m( 3v_0\frac{d^2x}{dt^2} + v_0t \frac{d^3x}{dt^3} +b\frac{d^3y}{dt^3}) \nonumber \\
  &=& \frac{m^2v_0^2t}{6((v_0t)^2+b^2)^{\frac{3}{2}}}.
\label{I_zz}
\end{eqnarray}
Introducing the notation $w=\frac{v_0t}{b}$, the triply differentiated  mass quadrupole for one black hole is:
\begin{eqnarray}
 I^{(3)}_{xx}&=& -\frac{m^2 v_0}{6b^2} \frac{w(2w^2 + 11)}{(w^2+1)^{\frac{5}{2}}}, \nonumber \\
 I^{(3)}_{yy}&=& \frac{m^2 v_0}{6b^2} \frac{w(w^2 + 10)}{(w^2+1)^{\frac{5}{2}}}, \nonumber  \\
 I^{(3)}_{zz}&=& \frac{m^2 v_0}{6b^2} \frac{w}{(w^2+1)^{\frac{3}{2}}}, \nonumber \\
 I^{(3)}_{xy}&=& \frac{m^2 v_0}{2b^2} \frac{w^2 - 2}{(w^2+1)^{\frac{5}{2}}},
\label{massQ}
\end{eqnarray}
and other components are zero. These mass quadrupoles are for one black hole only, 
so in work below we include the contribution of the second equal mass black hole, which doubles these moments.

We work out $^xH_{xx}$ explicitly to demonstrate the method introduced 
in Section $\S$ 2.2. We assume spin on only {\it one} black hole.
(To indicate the direction of the spin component 
generating the spin-quadrupole, $a_x$ in this example case, we include 
a leading label ${}^x$ on the symbol ${}^xH_{ij}$.) For the particle with velocity 
$v_x\approx +v_0$, we have:
\begin{eqnarray}
^xH_{xx}&=& a_x((x+\frac{m}{2})^2-\frac{1}{3}((x+\frac{m}{2})^2+y^2)) \nonumber \\
{}& &- a_x((x-\frac{m}{2})^2-\frac{1}{3}((x-\frac{m}{2})^2+y^2)), \nonumber \\ 
{}&=& \frac{2 a_x}{3} ((x+\frac{m}{2})^2 - (x-\frac{m}{2})^2),\nonumber \\
{ }&=& \frac{4 a_x m x}{3}. 
\end{eqnarray}
Similarly,
\begin{eqnarray}
^xH_{yy}&=& a_x(y^2-\frac{1}{3}((x+\frac{m}{2})^2+y^2)) \nonumber \\
{} & & - a_x(y^2-\frac{1}{3}((x-\frac{m}{2})^2+y^2)), \nonumber \\ 
{}&=&  -\frac{2 a_x m x}{3}\nonumber \\ 
{}&=&  ^xH_{zz},\nonumber \\
^xH_{xy}&=& a_x m y,
\end{eqnarray}
and others zero. (We used $z=0$.)
Also, 
\begin{eqnarray}
^yH_{xx}&=& a_x(x^2-\frac{1}{3}(x^2+(y+\frac{m}{2})^2)) \nonumber \\
{} & & - a_x(x^2-\frac{1}{3}((x^2+(y+\frac{m}{2})^2)) \nonumber \\ 
{}&=& -\frac{2a_ymy}{3}\nonumber \\
{}&=& ^yH_{zz},\nonumber \\
^yH_{yy}&=& \frac{4 a_y m y}{3},\nonumber \\
^yH_{xy}&=& a_y m x, 
\end{eqnarray}
and 
\begin{eqnarray}
^zH_{xz}&=& a_z m x,\nonumber\\
^zH_{yz}&=& a_z m y,
\end{eqnarray}
and others zero.

After differentiating and combining components, we have:
\begin{eqnarray}
H^{(3)}_{xx}&=&\frac{2m}{3}(2a_x\frac{d^3x}{dt^3}-a_y\frac{d^3y}{dt^3})\nonumber \\
 &= & \frac{m^2v_0(2a_x(2w^2-1)-3a_yw)}{6b^3(w^2+1)^{\frac{5}{2}}},\nonumber\\
H^{(3)}_{yy}&=&\frac{m^2v_0(-a_x(2w^2-1)+6a_yw)}{6b^3(w^2+1)^{\frac{5}{2}}},\nonumber\\
H^{(3)}_{zz}&=&-\frac{m^2v_0(a_x(2w^2-1)+3a_yw)}{6b^3(w^2+1)^{\frac{5}{2}}},\nonumber\\
H^{(3)}_{xy}&=&\frac{m^2v_0(3a_xw+a_y(2w^2-1))}{4b^3(w^2+1)^{\frac{5}{2}}},\nonumber\\
H^{(3)}_{xz}&=&\frac{m^2v_0a_z(2w^2-1)}{4b^3(w^2+1)^{\frac{5}{2}}},\nonumber\\
H^{(3)}_{yz}&=& \frac{3m^2v_0a_zw}{4b^3(w^2+1)^{\frac{5}{2}}}.
\label{spinQ}
\end{eqnarray}
The linear term (the first term) in Eq (\ref{kickEQ}) gives, for instance the force $dP_x/dt$ on the binary system:
\begin{eqnarray}
\frac{dP_x}{dt} &=& \frac{16}{45}(I^{(3)}_{xy}H^{(3)}_{xz}+(I^{(3)}_{yy}-I^{(3)}_{zz})H^{(3)}_{yz}). \label{dP_x/dt}
\end{eqnarray}
The force computed is 
applied to the total mass, so we find the 
individual black hole velocity by time integrating Eq (\ref{dP_x/dt}), using 
$dt= b/v_0$, and dividing the result by $2m$. The velocities are estimated in 
$km/s$ for $a_x\approx a_z\approx m$, $v \approx 1$, and $\frac{m}{b} \approx 
\frac{1}{2}$ (closest approach $= 4m$):
\begin{eqnarray}
 P_x &=& 2 \times (\frac{4}{45}\frac{m^4v_0^2}{b^5}a_z\frac{b}{v_0}\int_{-\infty}^{\infty}\frac{1}{(w^2+1)^3}dw) ~=~ 2 \times \frac{\pi}{30}(\frac{m}{b})^4a_zv_0, \\
v_x &\approx& 1962 ~km/s. \label{deltavx}
\end{eqnarray}
Similarly,
\begin{eqnarray}
\frac{dP_y}{dt} &=& \frac{16}{45}((I^{(3)}_{zz}-I^{(3)}_{xx})H^{(3)}_{xz}-I^{(3)}_{xy}H^{(3)}_{yz}), \nonumber \\
P_y &=& 2 \times (\frac{4}{45}\frac{m^4v_0^2}{b^5}a_z\frac{b}{v_0}\int_{-\infty}^{\infty}\frac{w}{(w^2+1)^3}dw), \nonumber \\
&=& 0; \\
\frac{dP_z}{dt} &=& \frac{16}{45}(I^{(3)}_{xx}-I^{(3)}_{yy})H^{(3)}_{xy}+I^{(3)}_{xy}(H^{(3)}_{yy}-H^{(3)}_{xx}), \nonumber \\
P_z &=& 2 \times (-\frac{2}{45}\frac{m^4v_0^2}{b^5}\frac{b}{v_0}\int_{-\infty}^{\infty}\frac{a_x(7w^2+4)}{(w^2+1)^4}+\frac{a_y(2w^3+5w)}{(w^2+1)^4}dw) \nonumber \\ 
&=& 2 \times (-\frac{3\pi}{40}(\frac{m}{b})^4a_xv_0), \\
v_z &\approx& -4415 ~km/s.\label{deltavz}
\end{eqnarray}
Odd integrands integrate to zero in our straight-line integration approximation. 
The accumulated $x-$velocity is the residual CM motion after the encounter, but it
is at most of order $10^{-2}$ of $v_0$, so may be unmeasureable.
However the $P_z$ estimate of Eq (\ref{deltavz}) is substantial. Notice that these estimated kicks 
are for {\it one} black hole with spin. If both are spinning, the symmetries of the equal mass orbit 
dictate that $a_i \rightarrow (a_1 -a_2)_i$. Hence equal magnitude oppositely directed spin doubles this kick velocity.

At this point we recall the limitations of these calculations, principally that the 
calculation of the dynamics is Newtonian. Our result is completely consistent and accurate 
in the Newtonian small-deflection limit, but the estimate Eq (\ref{deltavz}) is an extravagant 
extrapolation to $v_0=c$. In the absence of a General relativistic 2-body simulation, we can make 
only qualitative adaptations to relativity. One point to notice is that $\frac{m}{b}$ is half the 
deflection angle in the high-speed Newtonian limit. For a test body moving near $v=c$ past a central 
mass, in General Relativity the deflection at a given impact parameter and mass is twice the Newtonian 
result assuming $v=c$. This suggests that we might obtain the result estimated above from motion with 
twice the impact parameter.

\section{Quasi-Circular Inspiral}

To contrast our fly-by calculations above, we now calculate the kicks
when equal mass black holes ($m_{1}=m_{2}=m$) are in a circular 
orbit in the $x$-$y$ plane.  (In fact,  the loss of energy means the orbit spirals inword, so is only 
quasi-circular, but we assume a circular orbit, with the orbital separation an adjustably shrinking quantity
to mimic this energy loss.)  We choose the first black hole 
initially (t=0) at position $x = +d$, where the second black hole would be at
$x = -d$, with $2d$ as the ``circular orbit separation''. 
The third derivatives of the mass quadrupole components for just the first black hole are thus:
\begin{eqnarray}
 I^{(3)}_{xx}&=& 4md^2\omega^3\sin{2\omega t}, \nonumber \\
 I^{(3)}_{yy}&=& -4md^2\omega^3\sin{2\omega t}, \nonumber \\
 I^{(3)}_{xy}&=& -4md^2\omega^3\cos{2\omega t}.\label{circMQ} 
\end{eqnarray}
and other components are zero. The total differentiated mass quadrupole for the two equal-mass system is twice that given in Eqs (\ref{circMQ}).

The spin multipoles are calculated with the method described in $\S$ 2.2 so
that for one black hole, the two spin charges per component can be summed 
using the following coordinates:

For $a_x$:
\begin{eqnarray}
x_1&=&d\cos{\omega t} + (m/2),~~~y_1 ~=~ d\sin{\omega t},~~~z_1 ~=~ 0, \nonumber\\
x_2&=&d\cos{\omega t} - (m/2),~~~y_2 ~=~ d\sin{\omega t},~~~z_2 ~=~ 0.
\end{eqnarray}

For $a_y$:
\begin{eqnarray}
x_1 &=& d\cos{\omega t},~~~y_1 ~=~ d\sin{\omega t} + (m/2),~~~z_1 ~=~ 0, \nonumber\\
x_2 &=& d\cos{\omega t},~~~y_2 ~=~ d\sin{\omega t} - (m/2),~~~z_2 ~=~ 0.
\end{eqnarray}

For $a_z$:
\begin{eqnarray}
x_1 &=& d\cos{\omega t},~~~y_1 ~=~ d\sin{\omega t},~~~z_1 ~=~ + (m/2), \nonumber\\
x_2 &=& d\cos{\omega t},~~~y_2 ~=~ d\sin{\omega t},~~~z_2 ~=~ - (m/2).
\end{eqnarray}

Again, as in the fly-by case above, we use the fact that the spin is to 
lowest order parallel transported along the orbit, which is (quasi-) circular
here. The non-zero, third derivatives of the spin quadrupoles are thus:
\begin{eqnarray}
{}^{ x }H^{(3)}_{xx}  =  \frac{4}{3}a_xdm\omega^3\sin{\omega t},&~&~ {}^{ x }H^{(3)}_{yy}  =  -\frac{2}{3}a_xdm\omega^3\sin{\omega t}, \nonumber \\
{}^{ x }H^{(3)}_{zz}  =  -\frac{2}{3}a_xdm\omega^3\sin{\omega t},&~&~ {}^{ x }H^{(3)}_{xy}  =  -a_xdm\omega^3\cos{\omega t},\nonumber\\
{}^{ y }H^{(3)}_{xx}  =  \frac{2}{3}a_ydm\omega^3\cos{\omega t},&~&~ {}^{ y }H^{(3)}_{yy}  =  -\frac{4}{3}a_ydm\omega^3\cos{\omega t},\nonumber\\
{}^{ y }H^{(3)}_{zz}  =  \frac{2}{3}a_ydm\omega^3\cos{\omega t},&~&~ {}^{ y }H^{(3)}_{xy}  =  -a_ydm\omega^3\sin{\omega t},\nonumber\\
{}^{ z }H^{(3)}_{xz}  =  a_zdm\omega^3\sin{\omega t},&~&~ {}^{ z }H^{(3)}_{yz}  =  -a_zdm\omega^3\cos{\omega t},
\end{eqnarray}
and the non-zero, fourth derivatives of the spin octupoles, needed only for
the second term of Eq (\ref{kickEQ}), are:
\begin{eqnarray}
{}^{ x }H^{(4)}_{xxx}  =  \frac{72}{5}a_xd^2m\omega^4\cos{2\omega t},& & {}^{ x }H^{(4)}_{xyy}  =  -\frac{56}{5}a_xd^2m\omega^4\cos{2\omega t},~~~ {}^{ x }H^{(4)}_{xzz}  =  -\frac{16}{5}a_xd^2m\omega^4\cos{2\omega t},\nonumber\\
{}^{ x }H^{(4)}_{xxy}  =  \frac{64}{5}a_xd^2m\omega^4\sin{2\omega t},& & {}^{ x }H^{(4)}_{yyy}  =  -\frac{48}{5}a_xd^2m\omega^4\sin{2\omega t},~~~{}^{ x }H^{(4)}_{yzz}  =  -\frac{16}{5}a_xd^2m\omega^4\sin{2\omega t},\nonumber\\
{}^{ y }H^{(4)}_{xxx}  =  -\frac{48}{5}a_yd^2m\omega^4\sin{2\omega t},& & {}^{ y }H^{(4)}_{xyy}  =  \frac{64}{5}a_yd^2m\omega^4\sin{2\omega t},~~~~~{}^{ y }H^{(4)}_{xzz}  =  -\frac{16}{5}a_yd^2m\omega^4\sin{2\omega t},\nonumber\\
{}^{ y }H^{(4)}_{xxy}  =  \frac{56}{5}a_yd^2m\omega^4\cos{2\omega t},& & {}^{ y }H^{(4)}_{yyy}  =  -\frac{72}{5}a_yd^2m\omega^4\cos{2\omega t},~~~{}^{ y }H^{(4)}_{yzz}  =   \frac{16}{5}a_yd^2m\omega^4\cos{2\omega t},\nonumber\\
{}^{ z }H^{(4)}_{xyz}  =  8a_zd^2m\omega^4\sin{2\omega t},& & {}^{ z }H^{(4)}_{xxz}  =  8a_zd^2m\omega^4\cos{2\omega t},~~~~~~~{}^{ z }H^{(4)}_{yyz}  =  -8a_zd^2m\omega^4\cos{2\omega t}.~~~~~~~
\end{eqnarray}
Note that if both black holes have spin, ${}^cH^{(3)}_{ab}$ is computed by subtracting a similar formula for the second spin: ${}^cH^{(3)}_{ab} \propto (a_{c_1}-a_{c_2})$; and 
${}^cH^{(4)}_{abf} \propto (a_{c_1}+a_{c_2})$.

\subsection{First Term}

We now calculate the first term from Eq (\ref{kickEQ}),
\begin{eqnarray}
\label{kickEQ1}
\frac{dP_i}{dt}_{1^{st}} &=& \frac{16}{45}\epsilon_{ijk} I^{(3)}_{jl} H^{(3)}_{kl}\,.
\end{eqnarray}

$1^{st}$ term only:
\begin{eqnarray}
\frac{dP_x}{dt}_{1^{st}} &=& \frac{16}{45}(I^{(3)}_{xy}H^{(3)}_{xz}+I^{(3)}_{yy}H^{(3)}_{yz}), \nonumber \\
&=& 2 \times (\frac{64}{45}d^3m^2\omega^6a_z\sin{\omega t}), \nonumber \\
\frac{dP_y}{dt}_{1^{st}} &=& \frac{16}{45}(-I^{(3)}_{xx}H^{(3)}_{xz}-I^{(3)}_{xy}H^{(3)}_{yz}) \nonumber \\
&=& 2 \times (-\frac{64}{45}d^3m^2\omega^6a_z\cos{\omega t}), \nonumber \\
\frac{dP_z}{dt}_{1^{st}} &=& \frac{16}{45}(I^{(3)}_{xx}-I^{(3)}_{yy})H^{(3)}_{xy}+I^{(3)}_{xy}(H^{(3)}_{yy}-H^{(3)}_{xx}) \nonumber \\
&=& 2 \times (\frac{128}{45}d^3m^2\omega^6(a_y\cos{\omega t} - a_x\sin{\omega t})).
\label{P_1}
\end{eqnarray}
The first term of Eq (\ref{kickEQ}) is linear in spin,
consistent with computational simulations as seen in
\citet{2007ApJ...661..430H,2007PhRvD..76h4032H}. The `` 2 $\times$ ''
accounts for the two black holes of the system (doubling the mass quadrupole of a single 
black hole). 

For arbitrary orientation of spin 
of magnitude $a$, the components are simply $a_{x}= a\sin\theta\cos\varphi$, 
$a_{y}= a\sin\theta\sin\varphi$ and  $a_{z}= a\cos\theta$. 
Above, the $a_i$ are the spin components of just the one black hole that is spinning, but
if both holes were spinning, we replace $a$ with ($a_{1} - a_{2}$).

The circular orbit case presented here is based on Newtonian orbits, 
which specify frequency as a function of the Newtonian separation:

\begin{equation}
 \omega = \sqrt{\frac{m}{d^3}}.
\end{equation}

\subsection{Second Term}

With the symmetries of \citet{2007ApJ...661..430H,2007PhRvD..76h4032H}, the second, 
nonlinear term in Eq (\ref{kickEQ}) vanishes identically. However if the spins are not 
equal in magnitude or not anti-aligned, this nonlinear term does not
vanish, implying a quadratic contribution to kick velocity.

We calculate the second, quadratic, term from Eq (\ref{kickEQ}),
\begin{eqnarray}
\label{kickEQ2}
\frac{dP_i}{dt}_{2^{nd}} &=& \frac{4}{63} H^{(4)}_{ijk}
 H^{(3)}_{jk}\,.
\end{eqnarray}

$2^{nd}$ term only:
\begin{eqnarray}
\frac{dP_x}{dt}_{2^{nd}} &=& \frac{4}{63}(H^{(4)}_{xxx}H^{(3)}_{xx} + H^{(4)}_{xxy}H^{(3)}_{xy} + H^{(4)}_{xyy}H^{(3)}_{yy} + H^{(4)}_{xzz}H^{(3)}_{zz} + H^{(4)}_{xxz}H^{(3)}_{xz} + H^{(4)}_{xyz}H^{(3)}_{yz}) \nonumber \\
&=& \frac{16}{315} d^3 m^2 \omega^7 [- \sin{\omega t}(26a_{x}^{2}+23a_{y}^{2}+ 10a_{z}^{2}) - \sin{3 \omega t}(10a_{x}^{2} - 9a_y^2)\nonumber \\
&& ~~+~ a_x a_y (3 \cos{\omega t} + 11 \cos{3 \omega t})], \nonumber \\
\frac{dP_y}{dt}_{2^{nd}} &=& \frac{4}{63}(H^{(4)}_{xxy}H^{(3)}_{xx} + H^{(4)}_{xyy}H^{(3)}_{xy} + H^{(4)}_{yyy}H^{(3)}_{yy} + H^{(4)}_{yzz}H^{(3)}_{zz} + H^{(4)}_{yyz}H^{(3)}_{yz} + H^{(4)}_{xyz}H^{(3)}_{xz})\nonumber \\
&=& \frac{16}{315} d^3 m^2 \omega^7 [\cos{\omega t}(23a_x^2 + 26a_y^2 + 10a_z^2) + \cos{3 \omega t}(-9a_x^2 + 10a_y^2) \nonumber \\ 
&& ~~+~ a_xa_y(-3\sin{\omega t} + 11\sin{3 \omega t})], \nonumber \\
\frac{dP_z}{dt}_{2^{nd}} &=& \frac{4}{63}(H^{(4)}_{xzz}H^{(3)}_{xz} + H^{(4)}_{yzz}H^{(3)}_{yz} + H^{(4)}_{xxz}H^{(3)}_{xx} + H^{(4)}_{yyz}H^{(3)}_{yy} + H^{(4)}_{xyz}H^{(3)}_{xy}) \nonumber \\
&=& \frac{16}{315} d^3 m^2 \omega^7 a_z (11 a_y \cos{\omega t} + 5 a_y \cos{3 \omega t} - 11 a_x \sin{\omega t} + 5 a_x \sin{3 \omega t}).
\label{P_2}
\end{eqnarray}

\begin{figure}
\epsscale{.5}
\plotone{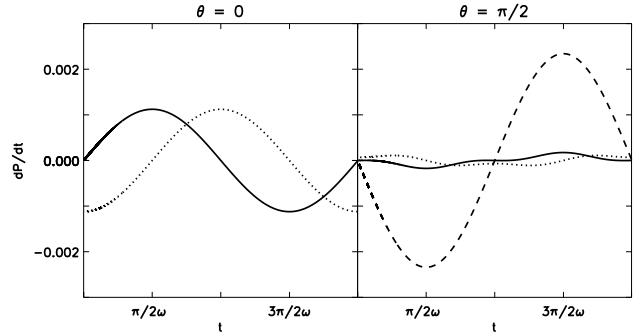}
\caption{\label{fig2} 
Quasi-circular kick-components when the spins ($a \sim 0.6m$) are perpendicular to the plane ($\theta=0$) and in the plane ($\theta=\frac{\pi}{2}$).
The $x$-component of the kick is the solid line, the $y$-component is the dotted line, and the $z$-component is the dashed line. Notice
the quadratic effects appearing when the spin is in the plane. Quadratic effects will become comparable to the linear effects only when $a \sim 1$.}
\end{figure}

\begin{figure}
\epsscale{.5}
\plotone{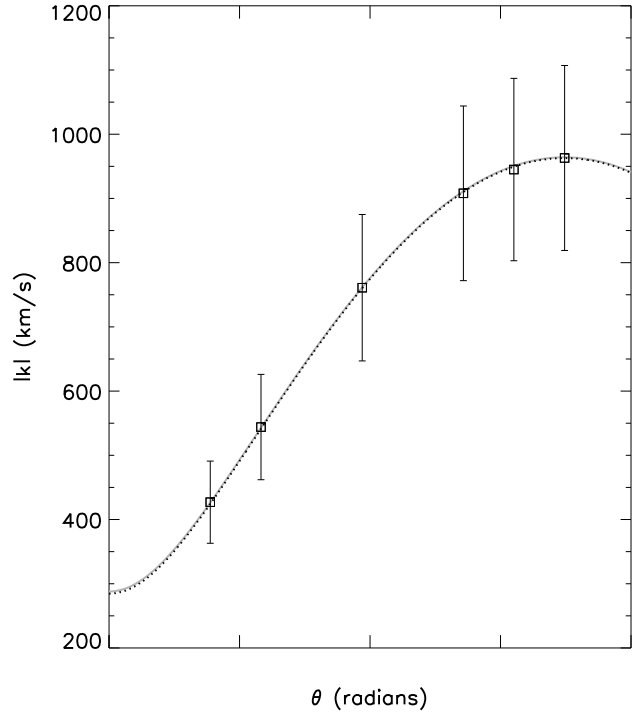}
\caption{\label{fig1} 
Comparing computational result for kick velocities (squares - the \citet{2007PhRvD..76h4032H} computational B-Series) as a function of spin,
to two analytic expressions:
Dotted line - \citet{2007arXiv0712.2819B} expansion; solid grey line - multipole 
analysis (this work). Because of our ambiguity in the ``last orbit" radius, and in the phase of 
the fraction of the ``last orbit" 
determining the kick, we fit two parameters: the maximum kick, and the phase.}
\end{figure}

The total force of the kick-components (first and second terms of Eq (\ref{kickEQ}) together)
can be compared in Fig (\ref{fig2}) for the case where only one of two equal mass black holes is
spinning at $a = 0.6m$ with a separation of $6m$. Notice that when the spin is perpendicular to the plane
($\theta=0$), the $z$-component of the kick vanishes. When the spin is in the plane
($\theta=\frac{\pi}{2}$), the linear term predicts only a $z-$ component to the kick. In fact 
the in-plane kick components ($x$ \& $y$) are nonzero due to the
non-linear contribution found in the second term of the multipole analysis formula (Eq (\ref{kickEQ})).

With any choice of the orbital phase and in any spin configuration the largest kicks are those 
linear in spin. The contributions quadratic in spin become comparable to the linear terms only for $a \sim1 $. 
However we are interested in the general solutions rather than simply the largest kicks, and 
there exists a range of spin angles where the quadratic behavior of the second term in Eq (\ref{kickEQ}) 
can be seen to dominate over the first, linear term in some components . This occurs when the spin 
angle $\theta$ satisfies $|\theta -\frac{\pi}{2}| < \frac{\pi}{12}$, that is, when the $x$-$y$ 
spin components (orbital plane) are much larger than the $z$ spin component. In this configuration the 
the $x$-$y$ component of the kick is dominated by the quadratic contribution, while the $z$-component 
of the kick is more than an order or magnitude larger. 

Without an analytical 2-body solution, there is substantial ambiguity in converting this to
a relativistic specification. An even more serious problem arises from the kick
formulae predicting forces ($dP_i/dt$) whose vertical component oscillates and whose in-plane
components rotate with the orbit. If the orbit were strictly circular, the average of the kick
would be zero, though the system would execute periodic motion due to the asymmetric radiation. 
In fact the orbit is only quasi-circular, and shrinks slowly due to gravitational radiation. 
Eventually the holes either spiral until they disappear behind a common black hole horizon, 
or enter a final plunge to the horizon. This sudden cutting off means the net kick
can be modelled by considering the ``last" (quasi-) circular fractional orbit.
Such a concept is ambiguous at best because no analytic prescription describes
the motion. However, numerical experiments do show kicks, and we can extract their 
dependence on orbital parameters. It is of great utility give simple analytical 
predictions, and thus we produce net, effective kicks depending on the phase of 
the orbit as it finally merges, and parametrized by the fraction of the ``last" orbit that 
contributes the net kick.

\section{Discussion}

It is of interest to compare the kicks from the quasicircular orbits, to the ones from the
hyperbolic flyby. For instance, Eq(20) for hyperbolic flyby has a kick velocity 
\begin{equation}
v_z \sim \frac{1}{4} (\frac{m}{b})^4 (\frac{a_x}{m})v_0.
\label{HY}
\end{equation}
To compare this to the circular orbit result, consider the force, Eq.(29):
\begin{equation}
\frac{dP_z}{dt} \sim 5 d^3 m^2 \omega^6 a_x \sim 5 m^2 a_x \frac{m^5}{d^6},
\end{equation}
where we used Eq(30).
The kick velocity is approximated by multiplying $\frac{dP_z}{dt} $ by a fraction ($\epsilon/2 \pi$) of an
orbital period, and dividing by $2m$. Thus:
\begin{equation}
v_z \sim \epsilon\frac{5}{2} (\frac{m}{d})^4 (\frac{a_x}{m})(\frac{m}{d^2})\sqrt{\frac{d^3}{m}}\sim \epsilon\frac{5}{2^{\frac{3}{2}}} (\frac{m}{d})^4 (\frac{a_x}{m})v_{orbit}.
\label{QC}
\end{equation}
The process ``multiplying $\frac{dP_z}{dt} $ by a fraction of an
orbital period" summarizes integrating the force for the relevant period at the final plunge, since circular orbits do not produce
a net kick.

Equations(\ref{HY}) and (\ref{QC}) are very similar, differing (aside from numerical factors) 
by $v_0$ in the hyperbolic case being replaced by $\epsilon v_{orbit}$, and 
the impact parameter by the orbital radius. The quantity $\epsilon$
is poorly defined, but is likely less than unity. Also, while $v_0$ can in principle be very 
close to unity, the orbital velocity for a given orbital radius will be much less than the 
flyby velocity with an impact parameter  equal to that orbital radius. This suggests that hyperbolic-orbit kicks can
in principle be larger (much larger) than quasicircular kicks.
Numerical studies (\citet{healyEtAl2008}) confirm these
large kicks predicted by the multipole analysis for the fly-by case. 

Recently, \citet{2007arXiv0712.2819B} presented a spin expansion in order to
understand final quantities of binary black hole mergers, such as mass, kick velocity,
and spin vector. They consider two Kerr black holes in quasicircular orbits and Taylor expand some final quantity
in terms of the spins. Symmetry arguments remove excessive independent terms 
at each order. \citet{2007arXiv0712.2819B} discover second and third order spin contributions that lie beyond the 
empirical fitting formulas which come from post-Newtonian, linear dependence fits from
simulations. We compare our results to their expansion for a numerical
black hole binary simulation in \citet{2007PhRvD..76h4032H}, referred to as the ``B-series'' 
(\S IV C of \citet{2007arXiv0712.2819B}). In this particular case, equal mass black 
holes have oppositely directed equal-magnitude spins ($a = 0.6m$) lying in the $x$-$z$ plane 
($\theta_1=$[0, $\pi$] while $\theta_2=$[$\pi$, $2\pi$] for $BH_1$ and $BH_2$, respectively).
Fig (\ref{fig1}) compares the Herrmann et al. numerically computed points 
with the Boyle \& Kesden expansion and our multipole analysis for the resultant kick magnitude.
Because of our ambiguity in the ``last orbit" radius, and in the phase of 
the fraction of the ``last orbit" 
determining the kick, we fit two parameters: the maximum kick, and the phase, 
and we plot the sum of Eq(29) and Eq(32). 
We find a tight agreement between these three methods, and the quadratic contribution in Boyle \& 
Kesden expansion can be quantitatively understood with the quadrupole and octupole of the binary
using the multipole formula.

\acknowledgments
This work was supported by NSF grant PHY-0354842 and NASA grant NNG 04GL37G.  
We thank Pablo Laguna for extensive communications.  
\pagebreak

\bibliography{references}

\begin{thebibliography}{58}
\expandafter\ifx\csname natexlab\endcsname\relax\def\natexlab#1{#1}\fi
\expandafter\ifx\csname href\endcsname\relax
  \def\href#1#2{}\fi
\expandafter\ifx\csname urllinklabel\endcsname\relax
  \def\urllinklabel{[LINK]}\fi
\expandafter\ifx\csname adsurllinklabel\endcsname\relax
  \def\adsurllinklabel{[ADS]}\fi

\bibitem[Blanchet et al.(1990)]{1990MNRAS.242..289B} Blanchet, L., Damour,
T., \& Schaefer, G.\ 1990, \mnras, 242, 289

\bibitem[Boyle \& Kesden(2007)]{2007arXiv0712.2819B} Boyle, L., 
\& Kesden, M.\ 2007, ArXiv e-prints, 712, arXiv:0712.2819 

\bibitem[Campanelli et al.(2007)]{2007PhRvD..75f4030C} Campanelli, M., Lousto, C.~O., 
Zlochower, Y., Krishnan, B.,\& Merritt, D.\ 2007, \prd, 75, 064030

\bibitem[di Matteo et al.(2008)]{diMa..2008} di Matteo, T.,
Colberg, J., Springel, V., Hernquist, L.,
\& Sijacki, D.\ 2008, \apj, 676, 33

\bibitem[{Gourgoulhon(2007)}]{EricGourgoulhon} Gourgoulhon, E., 2007,
submitted to Journal of Physics: Conference Series,
for the Proceedings of the VII Mexican School on Gravitation and
Mathematical Physics, held in Playa del Carmen,
Quintana Roo, Mexico, November 26 - December 2, 2006,
 (arXiv:0704.0149v2)

\bibitem[Hawking(1977)]{1977SciAm.236...34H} Hawking, S.~W.\ 1977,
Scientific American, 236, 34

\bibitem[Healy et al.(2008)]{healyEtAl2008}  Healy, 
J., Herrmann, F., Hinder, I., Shoemaker, D.~M., Laguna, P., 
\& Matzner, Richard~A.\ 2008 [arXiv:0807.3292]

\bibitem[Herrmann et al.(2007a)]{2007ApJ...661..430H} Herrmann, F., Hinder,
I., Shoemaker, D., Laguna, P., \& Matzner, R.~A.\ 2007a, \apj, 661, 430

\bibitem[Herrmann et al.(2007b)]{2007PhRvD..76h4032H} Herrmann, F., Hinder,
I., Shoemaker, D.~M., Laguna, P., \& Matzner, R.~A.\ 2007b, \prd, 76, 084032

\bibitem[Kidder(1995)]{PhysRevD.52.821} Kidder, L.~E. 1995, Phys. Rev. D, 52, 821

\bibitem[Komossa et al.(2008)]{2008ApJ...678L..81K} Komossa, S., Zhou,
 H., \& Lu, H.\ 2008, \apjl, 678, L81

\bibitem[Misner, Thorne, \& Wheeler(1973)]{MTW..1973} Mizner, C.~W., Thorne,
K.~S., Wheeler, J.~A.\ 1973, Gravitation, W.H. Freeman, New York

\bibitem[Murgia et al.(2001)]{2001A&A...380..102M} Murgia, M.,
 Parma, P., de Ruiter, H.~R., Bondi, M., Ekers, R.~D., Fanti, R., \& Fomalont,
 E.~B.\ 2001,\aap, 380, 102 

\bibitem[Oohara \& Nakamura(1989)]{1989PThPh..82..535O} Oohara, K., \& Nakamura, T.\
1989, Progress of Theoretical Physics, 82, 535

\bibitem[Pretorius(2005)]{PhysRevL.95.101} Pretorius, F., Phys. Rev. Lett.
95, 121101 [arXiv:gr-qc/0507014]

\bibitem[Schnittman et al.(2008)]{2008PhRvD..77d4031}	
Schnittman, Jeremy D., Buonanno, Alessandra, van Meter, James R., Baker, John G., Boggs, William D., 
Centrella, Joan, Kelly, Bernard J., McWilliams, Sean T., 2008
Physical Review D, 77, 044031

\bibitem[Shields \& Bonning(2008)]{2008arXiv0802.3873S} Shields, G.~A., \&
Bonning, E.~W.\ 2008, ArXiv e-prints, 802, arXiv:0802.3873

\bibitem[{Thorne(1980)}]{RevModPhys.52.299}
Thorne, K.~S. 1980, Rev. Mod. Phys., 52, 299

\bibitem[Thorne(1996)]{1996qpct.conf..101T} Thorne, K.~S.\ 1996, Quantum
Physics, Chaos Theory, and Cosmology, 101

\bibitem[Washik et al.(2008)]{matznerEtAl2008} Washik, M.~C., Healy, 
J., Herrmann, F., Hinder, I., Shoemaker, D.~M., Laguna, P., 
\& Matzner, R.~A.\ 2008, [arXiv:0802.2520] 

\bibitem[Whitaker \& van Dokkum(2008)]{Whit..2008} Whitaker, K.~E., \& van
Dokkum, P.~G.\ 2008, \apjl, 676, L105







\end{thebibliography}

\end{document}